\documentclass{aa}
\usepackage{epsfig}
\def\simle{\lower 2pt \hbox {$\buildrel < \over {\scriptstyle \sim }$}}
\def\simge{\lower 2pt \hbox {$\buildrel > \over {\scriptstyle \sim }$}}
\begin{document}
\thesaurus{02 
           (13.07.1;      %Gamma rays : bursts 
            08.19.1;      %Stars : statistics 
            09.03.2;      %ISM : cosmic rays 
            12.04.3)}      %Cosmology : distance scale 

\title{The jet-disk symbiosis model for Gamma Ray Bursts: cosmic ray 
and neutrino background contribution}
\author{G. Pugliese $^1$ \and H. Falcke $^1$ \and Y. P. Wang $^{2,3}$ \and 
P. L. Biermann $^{1,4}$}
\institute{$^1$ Max-Planck-Institut f\"{u}r Radioastronomie, Auf dem
H\"{u}gel 69, D-53121 Bonn, Germany \\ 
$^2$ Purple Mountain Observatory, Academica Sinica, Nanjing 210008, China \\ 
$^3$ National Astronomical Observatories, Chinese Academy of Sciences \\ 
$^4$ Department of Physics and Astronomy, University of Bonn, D-53121 Bonn, 
Germany}
\offprints{G. Pugliese, pugliese@mpifr-bonn.mpg.de}
\date{Received ??? , Accepted ???}
 \maketitle\markboth{Pugliese et al.: GRBs: CR and $\nu$'s contributions}
{Pugliese et al.: GRBs: CR and $\nu$'s contributions} 

\begin{abstract}

The relation between the cosmological evolution of the jet-disk symbiosis 
model for GRBs and the cosmic rays energy distribution is presented. We 
used two different Star Formation Rates (SFR) as a function of redshift 
and a Luminosity Function (LF) distribution to obtain the distribution in 
fluence of GRBs in our model and compare it with the data. We show a good 
agreement between the fluence distribution we obtain and the corrected data 
for the 4B BATSE catalogue. The results we obtain are generally valid for 
models that use jet physics to explain GRB properties. 

The fluence in the gamma ray band has been used to calculate the energy in 
cosmic rays both in our Galaxy and at extragalactic distances as a function 
of the redshift. This energy input has been compared with the Galactic and 
extragalactic spectrum of cosmic rays and neutrinos. Using our jet disk 
symbiosis model, we found that in both cases GRBs cannot give any significant 
contribution to cosmic rays. We also estimate the neutrino background, 
obtaining a very low predicted flux. We also show that the fit of our model
with the corrected fluence distribution of GRBs gives strong constraints of 
the star formation rate as a function of the redshift. 

\end{abstract}

\section{Introduction}

More than 30 years after their di\-sco\-ve\-ry, thanks to the Burst and 
Transient Source Experiment (BATSE) and the Italian-Dutch satellite 
BeppoSAX, the scientific community now knows that Gamma Ray Bursts (GRBs) 
are isotropically distributed in the sky (Fishman $\&$ Meegan 1995) and that 
at least some of them are at cosmological distances (GRB970228: Djorgovski 
et al. 1999b, GRB970508: Metzger et al. 1997, GRB971214: Kul\-kar\-ni 
et al. 1998, GRB980613: Djorgovski et al. 1999a, GRB980703: Djorgovski et 
al. 1998, GRB990123: Hjorth et al. 1999, GRB990510: Vreeswijk et al. 
1999, GRB990712: Galama et al. 1999). But the present data available for 
redshift position and host galaxy localization are still too few to give us 
good statistics to study the evolution of GRBs and their redshift distribution. 
Before the discovery of GRB afterglows by BeppoSAX, the only way to study 
their distributions was to compare some GRB properties (like for example the 
intensity), with some parametric models (Fenimore $\&$ Bloom 1995, Cohen $\&$ 
Piran 1995, Kommers et al. 1999). Because of this lack of information, it is 
still necessary to assume that GRBs follow the statistical distribution of 
some other better known objects to obtain the GRBs fluence or flux distribution 
itself. 

The origin of GRBs is still controversial. According to different models, 
their progenitor can be identified with the merging of two neutron stars, or 
with the collapse of a massive star. In the model presented by Pugliese et al. 
(1999), GRBs are created inside a pre-existing jet in a binary system formed 
by a neutron star and an O/B/WR companion, where the input energy comes from 
the collapse of the neutron star into a black hole and the emission is due to 
synchrotron radiation from the ultrarelativistic shock waves that propagate 
along the jet with a low-energy cut-off in the electron distribution. 
Following this scenario, the birth 
of GRBs cannot happen too far from the region where the progenitor formed, 
and this implies that their rate should be connected with the Star Formation 
Rate (SFR). Already other authors studied the connection between the SFR and 
GRBs flux distribution. For example, Wijers et al. (1998) showed that the 
assumption that the GRB rate is proportional to the SFR in the universe is 
consistent with the GRB flux distribution. 

In Sect. 2 we calculate the cumulative distribution of GRB fluences 
using two SFR distributions as a function of redshift, 
the one by Miyaji (Miyaji et al. 1998), and the other by Madau (Madau et 
al. 1996). We compared it with the data from the BATSE catalogue. In Sect. 
3 we calculate the maximum energy available in our model to obtain high energy 
cosmic rays. In Sect. 4 we present our results for the contribution of 
GRBs to the cosmic rays distribution, both Galactic and extragalactic and in 
Sect. 5 the eventual contribution from GRBs to the neutrino flux. 

\section{The rate of GRBs}

The aim of this section is the calculation of the number of GRBs per year 
and per 100 $\rm Mpc^3$ as a function of redshift, assuming beamed emission 
in a jet, (e.g. Pugliese et al., 1999). The following calculations are quite 
general and we emphasize that their validity does not depend on the particular 
model used. 

To obtain a result as close as possible to the data, we chose the fluence as 
the quantity that can well represent the characteristics of GRBs. We follow 
Petrosian $\&$ Lloyd (1997), who showed that the fluence is the most appropriate 
parameter to study the cosmological evolution of GRBs. The relation between 
the redshift $z$ and the fluence $f$ is given by (Petrosian $\&$ Lloyd
1997) as a function of the specific luminosity L expressed in 
$[\rm {erg \; s^{-1} \; Hz^{-1}}]$: 

\begin{equation}
f = {{L \, \Delta t \, \Delta \nu} \over {4 \pi d^2_{\rm L} (1 + z)^{{\rm 
\alpha} -3}}} \quad {\rm{erg/cm^{2}}}. 
\label{plato}
\end{equation}
Here $\alpha$ is the photon flux spectral index, $d_{\rm L} = (2c /h) 
(1 + z - \sqrt{1 + z})$ is the cosmological luminosity distance (Weinberg, 
1972), and $H_0 = h \, (100 \; {\rm km \; s^{-1} \; Mpc^{-1}})$ is 
the Hubble constant. In the transformation from the emitter frame to the 
observer frame the two corresponding redshift contributions from the 
frequency and the time dependence are cancelled. 

In our model (Pugliese et al. 1999), we used $\alpha = 2$, and the 
corresponding value for the fluence is: 

\begin{equation}
%f = {{L({\rm erg/s}) (1 + z)} \over {4 \pi d^2_{\rm L}}} \quad 
f = {{L \, \Delta t \, \Delta \nu} \over {4 \pi d^2_{\rm L}}} (1 + z) 
\quad {\rm{erg/cm^{2}}}. 
\label{aristotele}
\end{equation}

In this way the redshift as a function of the fluence is: 

\begin{equation}
1 + z(f,f_{\star}) = {\biggl( {1 \over 2} \sqrt {f_{\star} \over f} + 1 
\biggr)}^2, 
\label{socrate}
\end{equation}
where $f_{\star} = {{L(\nu_1) \Delta t \Delta \nu} \over {4 \pi^2}} 
{h^2 \over c^2}$ $\rm {erg/cm^2}$ is the reference fluence and $L(\nu_1)$ 
is expressed in $[\rm {erg \; s^{-1} \; Hz^{-1}}]$. 

At this point of our calculations it is important to define the role of the 
parameter $f_{\star}$. In fact a relevant question is whether GRBs are 
standard candles (i.e. $f_{\star}$ is constant) or whether they are 
distributed with a Luminosity Function (LF) with, e.g., a power law in 
$f_{\star}$. Developing our calculations, we arrived at the same results 
found by Kommers et al. (1999). Here the authors showed that the best model 
is the one in which the star formation rate is combined with the luminosity 
function distribution. In agreement with them, also in our model GRBs cannot 
be considered as standard candles. We use a distribution for the luminosity 
function with a power law index $-\beta$, and the corresponding law for 
the parameter $f_{\star}$ is given by: 

\begin{equation} 
\phi(f_{\star}) d f_{\star} =  \phi_0 f_{\star}^{- \beta} \; d f_{\star}, 
\label{pitagora}
\end{equation}
where $\phi_0$ is the normalization parameter equal to $1/(f_{\star 
\rm{b}}^{-1} - f_{\star \rm{a}}^{-1})$. We adopt $f_{\star \rm{a}} = 
10^{-8} \rm{erg/cm^2}$, and $f_{\star \rm{b}} = 2.2 \times 10^{-4} 
\rm{erg/cm^2}$. 

Here we do not yet answer the question what the physics of this luminosity 
function may be. It is plausible in the context of our model, that it is 
directly connected to the mass flow in the pre-existing jet prior to the 
GRB explosion. If this were the correct interpretation, the mass accretion 
rate, and correspondingly, the mass flow rate in the jet may follow a power 
law distribution, a point which we will pursue elsewhere. 

We also calculate the GRB rate using two different star formation rates 
and compare the corresponding results with the data. 

\subsection{The number count}

The number count of GRBs sources is given by the expression $dN(z) = 
F(z) \, (dt/dz) \, dV \, dz$, where $F(z)$ is equal to the product 
of the SFR $\psi(z)$ and the luminosity function $\phi(z)$. 
Obviously the luminosity function may change with redshift $z$, but 
for simplicity and as a first step, we use this ansatz. In term of 
the fluence $f$ of GRBs in the gamma ray band, and of the reference 
fluence $f_{\star}$, the number count is: 

\begin{eqnarray}
dN(f,f_{\star}) 
& = & [\psi (f,f_{\star})] {\biggl[ {{dt} \over {d z}} 
      (f,f_{\star}) \biggr]} {\biggl[ 4 \pi d^2_{\rm L}(f,f_{\star}) 
      {{d \, d_{\rm L}} \over {d z}} (f,f_{\star}) \biggr]} 
      \nonumber \\
&   & \times [d z(f,f_{\star})], 
\label{epicuro}
\end{eqnarray}
where: 

\begin{description}

\item[$\bullet$]

The first term $\psi(z(f,f_{\star}))$ represents the star formation rate 
as a function of the redshift. 

\item[$\bullet$]

The second term represents the temporal interval in which the rate is 
calculated. It is given by: 

\begin{equation}
{{d t} \over {d z}}(f,f_{\star}) = -{1 \over h} {1 \over {({1 \over 2} 
\sqrt {f_{\star}/f} + 1)}^5}. 
\label{catullo}
\end{equation}

\item[$\bullet$]

The third term is the interval of volume as a function of the fluence. 
It corresponds to the value: 

\begin{equation}
4 \pi d^2_{\rm L} {d \over {d z}}d_{\rm L} = 4 \pi {c^3 \over h^3} 
{f_{\star} \over f} {\biggl( {1 \over 2} \sqrt {f_{\star} \over f} 
+ 1 \biggr)} {\biggl( \sqrt {f_{\star} \over f} + 1 \biggr)}. 
\label{lucrezio}
\end{equation}

\item[$\bullet$]

The fourth term represents the redshift interval and it is given by: 

\begin{equation}
d z(f,f_{\star}) = - {1 \over {2 \pi}} {{\sqrt f_{\star}} \over 
{(\sqrt f)}^3} {\biggl( {1 \over 2} \sqrt {f_{\star} \over f} + 1 
\biggr)} \; d f. 
\label{cesare}
\end{equation}

\end{description}

The integration over the parameter $f_{\star}$ will be shown in the 
Eq. 11. 

\subsection{The GRB rate distribution using the Miyaji SFR}

Here we use the star formation rate as a function of the redshift presented 
by Miyaji et al. (1998), and compare the corresponding GRB rate with the data 
corrected for selection and calibration effects by Petrosian (priv. comm.) for 
the 4B BATSE catalogue. 

In their paper Miyaji et al. (1998), calculated the distribution of Seyfert 
galaxies as a function of redshift. We considered this same distribution for 
the SFR, and approximated it with the following function: 

\begin{equation}
\psi(z) = A \; {\rm{exp}} (a \; z), 
\label{seneca}
\end{equation}
where $z$ is the redshift, $A \simeq 10^{-6} [h^3 \; \rm{Mpc^{-3}}]$, 
$a=2.5$, and we used $\Omega=1$, $\Lambda=0$ as a simple reference. This law 
is valid up to $z=2$, and then continues as a constant to high redshifts. 

\begin{figure}[h]
\vspace{6.4cm}
\includegraphics{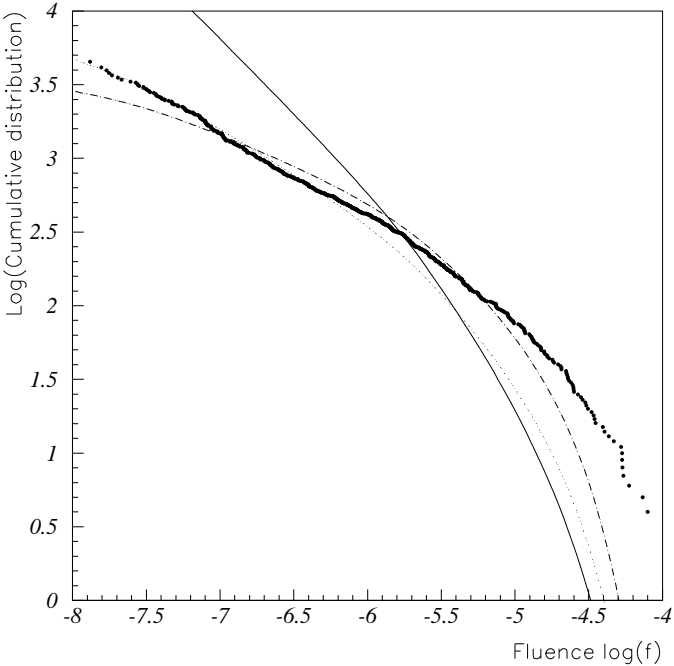} 
\vspace{2.4cm}
\caption[]{\label{rawdata} 
The corrected data (Petrosian priv. comm.) from the 4B BATSE catalogue 
(full circles) are compared with the distribution of GRB fluence 
calculated using the Miyaji SFR and a power law luminosity function 
index $\beta=1.5$ (dotted line), $\beta=2.0$ (dashed line), and 
$\beta=3.0$ (solid line).} 
\end{figure}

This SFR as a function of the fluence is given by: 

\begin{equation}
\psi(f,f_{\star}) = A \; {\rm{exp}} {\biggl[ a {\biggl[ {\biggl( {1 \over 2} 
\sqrt {f_{\star} \over f} + 1 \biggr)}^2 - 1 \biggr] }\biggr] }. 
\label{svetonio}
\end{equation}

Using the LF given in Eq.~\ref{pitagora}, the rate of GRB fluence is obtained 
from the product of the terms of the Eqs.~\ref{catullo} -~\ref{cesare}, 
and Eq.~\ref{svetonio}, where we now integrate over $f_{\star}$, that is: 

\begin{eqnarray}
N (f) \; df 
& = & 2 {c^3 \over h^4} A \; \phi_0 (f^{-\beta} \; df) 
      \int^{x_{\rm{max}}}_{x_{\rm{min}}} x^{-1/2} \times \nonumber \\ 
&   & {\rm{exp}} {\biggl[ {\rm{a} {\biggl[ {\biggl( {1 \over 2} \sqrt {x} 
      + 1 \biggr)}^2 - 1 \biggr] }}\biggr] } \times \nonumber \\ 
&   & {{\biggl( {1 \over 2} \sqrt {x} + 1 \biggr)}} {\biggl( \sqrt 
      {x} + 1 \biggr)}^{-3} \; d x, 
\label{cicero}
\end{eqnarray}
where $x = {f_{\star}/f}$, and $x_{\rm{max}}$ and $x_{\rm{min}}$ are 
the limits of integration defined as the maximum and the minimum of the 
intersection between the interval $[f_{\star \rm{a}},f_{\star \rm{b}}]$ 
relative to the LF distribution and the interval in which the ratio 
$f_{\star}/f$ is defined. 

In Fig.~\ref{rawdata} we plotted the cumulative curves corresponding to 
different luminosity function power law indexes ($\beta=1.5$ dotted line, 
$\beta=2.0$ dashed line, and $\beta=3.0$ solid line) in Eq.~\ref{cicero} 
and compared them with the corrected data of the 4B catalogue from BATSE. 
It is evident that even if we change the luminosity function distribution 
index, we cannot obtain any better fit or calculate any 
GRB rate using this SFR distribution. 

\subsection{The GRB rate distribution using the Madau SFR}

In this paragraph we calculate the GRB rate based on the same procedure as 
in the last section, using the SFR as a function of redshift presented by 
Madau et al. (1996) and the luminosity function distribution of 
Eq.~\ref{pitagora}. 

\begin{figure}[h]
\vspace{0.2cm}
\epsfig{file=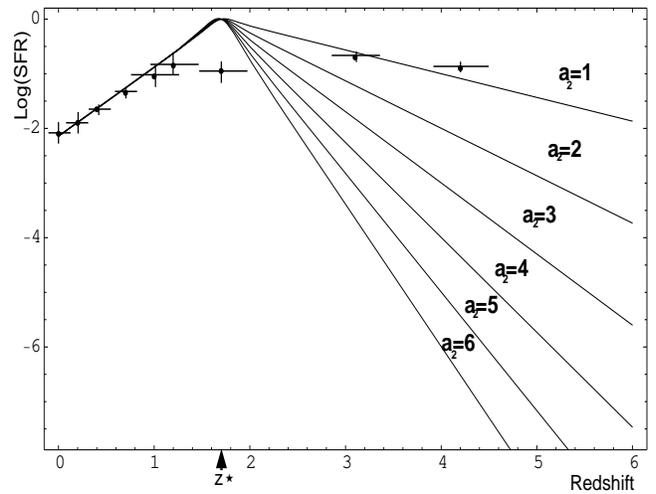,scale=2.0,height=6.5cm, 
        width=8.5cm,angle=0}
\vspace{0.2cm}
\caption[]{\label{madau} 
Madau SFR curves at different slopes after the redshift 
$z^{\star}=1.7$. The full circles are the data points 
from Steidel (1999). In this article, Steidel considers the 
curve $a_2=1$ as one of the many possible curves consistent 
with both the current data on the far-IR background and the 
galaxies detected in the new submillimiter band.} 
\end{figure}

We approximated the Madau SFR by the combination of two exponential functions. 
The first part of the curve is: 

\begin{equation}
k_1(z) \simeq {\rm{exp}} (a_1 \; z), 
\label{seneca}
\end{equation}
where $z$ is the redshift, and $a_1=2.9$ is a constant given by the IR 
source counts. This law is valid up to $z^{\star}=1.7$, after which it 
is substituted by the function:  

\begin{equation}
k_2(z) \simeq {\rm{exp}} (-a_2 \; z), 
\label{plinio}
\end{equation}
and $a_2$ is a constant. 

Because of the uncertainties of the dust extinction at high redshift and also 
the difficulties in the redshift determination for the SCUBA sources (Sanders 
1999), the star formation history beyond the redshift $z=2$ is still unclear, 
therefore we considered different slopes for this part of the Madau SFR curve 
and plotted them in Fig.~\ref{madau} together with the experimental data 
given by Steidel (1999). 

The cumulative GRB fluence distribution is obtained from Eq.~\ref{cicero}, 
where the SFR function is substituted with Eq.~\ref{seneca} and 
Eq.~\ref{plinio}. The parameters that we can change to fit the data are the 
redshift $z^{\star}$, the power law index $a_2$ in the Eq.~\ref{plinio} and 
the power law index $\beta$ and the upper limit $f_{\star \rm{b}}$ of the 
interval in which the luminosity function distribution is defined. Following 
one of the possible curves that fit the data shown by Steidel (1999), we chose 
$z^{\star}=1.7$, $\beta$ determines the slope of the cumulative distribution 
curve, $a_2$ defines curves with different slope, and variations in 
$f_{\star \rm{b}}$ correspond to little changes in the part of the curve 
relative to the strongest GRBs, we chose $f_{\star \rm{b}}=2.2 \times 10^{-4} 
\rm{erg/cm^2}$. To reproduce the corrected data by Petrosian of the 4B 
catalogue, we probed all the different cases changing the slopes of the SFR 
and the power law index of the fluence distribution. The only values for the 
power law index of the luminosity distribution that successfully fits the data 
was $\beta=1.55$. 

\begin{figure}[h]
\vspace{0.2cm}
\epsfig{file=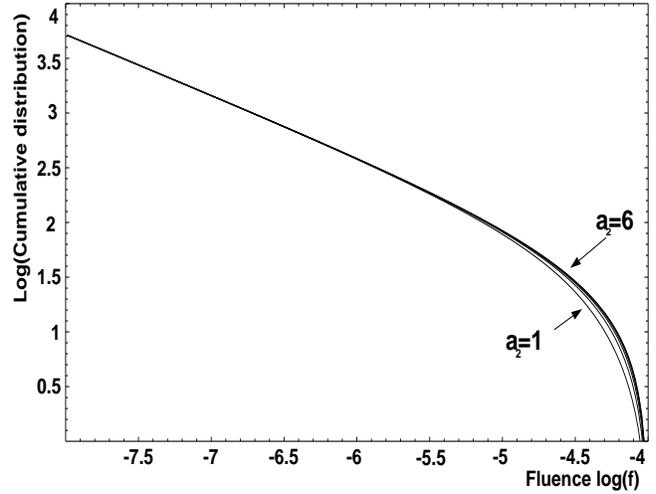,scale=2.0,height=6.5cm, 
        width=8.5cm,angle=0}
\vspace{0.2cm}
\caption[]{\label{cumul} 
Cumulative curves corresponding to the different slopes of 
Fig.~\ref{madau}.} 
\end{figure}

In Fig.~\ref{cumul} we show the cumulative fluence distribution corresponding 
to the different slopes of Fig.~\ref{madau} and the power law index $\beta = 
1.55$. We compared these curves with the data corrected for selection and 
calibration effects by Petrosian for the 4B BATSE catalogue: we can fit these 
data only if $a_2$ is in the range $0.8 \div 1.3$. In Fig.~\ref{corredata} 
the data are fitted with $a_2=1.0$. 

Considering the total number of GRBs in the BATSE catalogue, an observing time 
of 8 years, a volume scale of $h^{-3} 10^{10.8} \rm {Mpc^3}$, a beaming 
factor ${{4 \pi} \over {2 \pi \theta^2}} = 200 \, \theta^{-2}_{-1 \rm j}$, 
where $\theta_{-1 \rm j} = \theta_{\rm j} / (10^{-1} \rm{rad})$ is the 
opening angle of the jet, and the factor 22 coming from the integral of the 
distribution in Eq.~\ref{svetonio} calculated with these new SFR and LF, 
we obtain the following rate of GRBs: 

\begin{figure}[h]
\vspace{0.2cm}
\epsfig{file=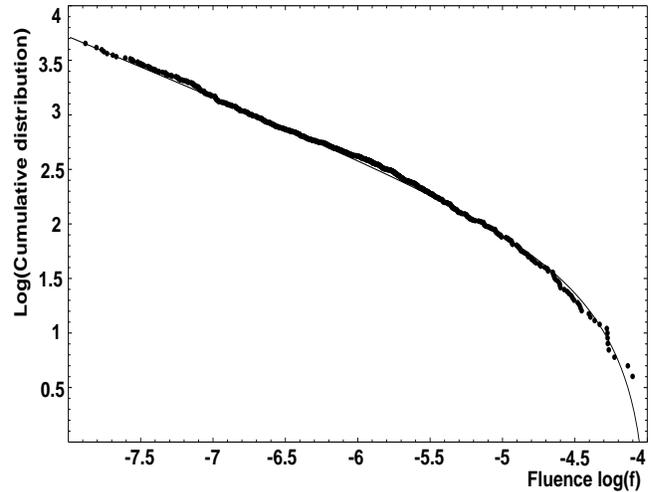,
        scale=4.0,height=6.5cm,width=8.5cm,angle=0}
\vspace{0.2cm}
\caption[]{\label{corredata} 
Comparison between the distribution of GRB fluence using 
the power law Luminosity function $\phi(f_{\star}) \propto 
f_{\star}^{-1.55}$, and the Madau SFR (solid line) and 
the 4B corrected data from Petrosian (priv.  comm.) (full circles).} 
\end{figure}

\begin{eqnarray}
10^{-5.4} (h^3 \theta_{-1 \rm{j}}^{-2}) \; \; {\rm {GRBs \; {(yr)}^{-1} \; 
(100 \; Mpc^{-3})}}. 
\label{ottaviano}
\end{eqnarray}

\vspace{0.4cm}

This value obtained considering beaming effects is close to the number 
given by Piran (1999). 

The result depends strongly on the power index, while it is not influenced 
by the interval $[f_{\star \rm{a}},f_{\star \rm{b}}]$ of the LF distribution 
we used.  We defined a lower limit for the fluence equal to $10^{-8} 
\rm{erg/cm^2}$, and a ratio $f_{\star \rm{b}}/f_{\star \rm{a}}$ equal to 
$10^4$. A change in the upper limit of this interval corresponds to a small 
change in the tail of the fluence distribution curve, for $x > 10^{-4}$. In 
our model, this interval in the fluence corresponds to an interval in the 
initial energy deposited in the jet in the range $[10^{48},10^{52}]$ ergs. 

\section{Maximum energy available}

Another important step is to check if it is possible to produce neutrinos and 
high energy cosmic rays with our model (Pugliese et al. 1999). Therefore we 
calculated the maximum energy available in the emission region. Irrespective 
of any acceleration mechanism, the maximum energy of charged particles, here 
protons, is that given by the spatial limit in the comoving frame, where the 
gyromotion just fits the space available: 

\begin{eqnarray} 
E_{\rm{max}}^{(\rm{ob})} = {1 \over 2} \gamma_{\rm{sh}} e \; {(B_{\rm j} 
\; \Delta z)}^{(\rm{sh})}. 
\end{eqnarray}
Here $e$ is the charge of an electron, $B$ is the magnetic field in the shock 
frame given by $B_{\rm j}^{\rm {(sf)}}(t) = 10.24 \, (E_{51}^{-1/4} \dot M_{-5 
\rm j}^{3/4} v_{0.3}^{-3/4}) (\epsilon^{-1/2} \theta_{-1 \rm j}^{-1} 
\gamma_{{\rm m},2}^{1/2}) \, t_5^{-3/4}$, and $\Delta z$ is the thickness of 
the emission region. For the calculation of the thickness we have two 
possibilities, and each of them corresponds to one of the following cases: 

\begin{description} 

\item [$\bullet$]
For the first case we used the thickness of the emission region in the shock 
frame (sh), given by $z/(4 \gamma_{\rm{sh}})$: 

\begin{eqnarray}
E_{\rm{max}}^{\rm{I (ob)}} 
& = & \gamma_{\rm{sh}} {\biggl(e \; B^{\rm{(sh)}} {z_{\rm j} \over 
      {4 \gamma_{\rm{sh}}}} \biggr)} \simeq 1.91 \times 10^{21} 
      (E_{51}^{1/4} \dot M_{-5{\rm j}}^{1/4} \times \nonumber \\ 
&   & v_{0.3}^{-1/4}) (\epsilon_0^{-1/2} \theta_{-1 {\rm j}}^{-1} 
      \gamma_{{\rm m},2}^{1/2}) \; t^{-1/4} \qquad {\rm{eV}}. 
\end{eqnarray}

\item [$\bullet$]
In the second case we considered the width of the shell: 

\begin{eqnarray}
E_{\rm{max}}^{\rm{II (ob)}} 
& = & \gamma_{\rm{sh}} (e \; B^{\rm{(sh)}} \theta_{\rm j} z_{\rm j}) 
      \simeq 0.93 \times 10^{23} E_{51}^{1/2} \times \nonumber \\ 
&   & (\epsilon_0^{-1/2} \gamma_{m,2}^{1/2}) \; t^{-1/2} \qquad {\rm{eV}}. 
\end{eqnarray}

\end{description}

At this point it is necessary to check when the thickness of the emission 
region is lower than the width, which corresponds to $E_{\rm{max}}^{\rm{I 
(ob)}} < E_{\rm{max}}^{\rm{II (ob)}}$. To do this we calculated the time at 
which these two quantities are equal: 

\begin{eqnarray}
t_{(E_{\rm{max}}^{\rm{I (ob)}} = E_{\rm{max}}^{\rm{II (ob)}})} \simeq 
5.62 \times 10^6 (E_{51}^{1/4} \dot M_{-5 \rm j}^{-1/4} v_{0.3}^{1/4}) 
\theta_{-1 \rm j} \quad {\rm{s}}. 
\end{eqnarray}

It means that for about two months $E_{\rm{max}}^{\rm{I (ob)}} 
< E_{\rm{max}}^{\rm{II (ob)}}$. Obviously, only $E_{\rm{max}}^{\rm{I (ob)}}$ 
is valid, so after 10 seconds the upper limit of the maximum energy available 
in our model is about $1.07 \times 10^{21} \, \rm{eV}$, while after two months 
the energy to consider is $E_{\rm{max}}^{\rm{II (ob)}}$. Adiabatic losses will 
diminish these energies for charged particles like protons. 

\section{Cosmic ray contribution} 

Cosmic rays are ionized nuclei, mainly protons, that extend from low 
energies (few hundred MeV) up to very high energies (about $3 \times 10^{20}$ 
eV). Their spectrum is described by a power law $(d N/d E) = E^{-({\rm \kappa} 
+1)}$, and shows two breaks in the slope. From low energies up to about $5 
\times 10^{15}$ eV, known as the knee, the spectrum follows a pure power law 
with $\kappa \simeq 1.7$. The detailed shape of this break and the precise 
position are still unknown. Beyond the knee up to about a second break point 
at $3 \times 10^{18}$ eV, known as the ankle, the pure power law has an 
index $\kappa \simeq 2$. 

It has been proposed (Biermann 1993, Stanev et al. 1993), that three 
components contribute to the cosmic ray spectrum: a) explosion of supernovae 
into a homogeneous interstellar medium (ISM), accelerating particles up to 
energies of about $10^5$ GeV. The spectrum for these particles is a power law 
with an index of -2.75, after considering the leakage from our Galaxy. b) 
Explosion of stars into their former stellar wind (like Wolf Rayet stars), 
producing particles with energies up to about $3 \times 10^9$ GeV. The 
corresponding spectrum switches at the knee from -2.67 to -3.07 and this 
difference in the spectral index derives from a diminution of the particle 
curvature drift energy gain. In Biermann's cosmic ray model for the Galactic 
component (Biermann 1997), the energetic protons are produced in the shocks 
of supernova explosions in the interstellar medium, while all the heavier 
elements are produced in the shock waves propagating in the stellar wind of 
the progenitor star. c) Production of particles with energies up to $10^{12}$ 
GeV from the hot spots of Fanaroff Riley class II radio galaxies. Their 
spectrum has an index -2 at the source, and one needs to take the interaction 
with the cosmological microwave background into account. 

We expect the spectrum of cosmic rays above about $5 \times 10^{19}$ GeV to 
be strongly attenuated because of the interaction of nuclei and protons with 
the 2.7 K cosmic microwave background, giving rise to the so-called 
Greisen-Zatsepin-Kuzmin (GZK) cut-off. The extragalactic sources cannot 
produce all the total cosmic ray energy density observed at Earth, 
but they could give a contribution to the ultra-high energy (UHE) part of 
the spectrum. The origin of the cosmic rays above $3 \times 10^{18}$ eV is 
not yet clear, but the common idea is that they are extragalactic and probably 
connected with the most powerful radio galaxies (Biermann $\&$ Strittmatter 
1987, Berezinsky $\&$ Grigor'eva 1988, Rachen $\&$ Biermann 1993, Rachen et al. 
1993). At the moment, our knowledge of the sources of the highest energy 
cosmic rays is limited by the small number of events detected by the present 
experiments. Considering that at $10^{20}$ eV the rate of cosmic rays is about 
1 event per $\rm{km}^2$ per century, it is clear that to detect them it is 
necessary to have both large aperture detectors and a long exposure time. 

In this general context, it is interesting to check the eventual Galactic 
and extragalactic energetic contribution given by GRBs to the cosmic ray 
spectrum in our model. In fact GRBs seem to be very powerful explosions 
and they inject a large amount of energy and elementary particles into the 
interstellar medium. 

We used the GRB rate obtained with the SFR from Madau and two different 
energetic approaches to calculate the contribution from GRBs to the cosmic 
rays and the neutrino spectra. First we considered that each GRB gives the 
same contribution equal to $10 \%$ of a fixed initial energy of $10^{51} \, 
\rm{ergs}$. Secondly we assume that each GRB contributes proportional to 
its own fluence, which we assume is distributed with a power law, 
corresponding to a range of initial energy of $[10^{48},10^{52}]$ ergs. 

\subsection{Extragalactic contribution}

For GRBs it is important to identify which particles contribute to the 
cosmic ray flux. As Rachen $\&$ M{\'e}sz{\'a}ros (1998) showed, during the 
main burst protons lose most of their energy because of adiabatic expansion, 
while neutrons can be better candidates to obtain ultra high energy cosmic 
rays (UHECR) and neutrinos. In fact neutrons carry about $80 \%$ of the 
proton energy, and because they are not coupled to the magnetic field, they 
can escape the fireball and through the $\beta$-decay give a cosmic ray 
proton spectrum. We followed this same logic in our calculations below. 

\begin{description} 

\item [a)]

We assumed that the total energy discharged by each GRB in the ISM for 
hadrons as well as neutrinos is equal to $10 \%$ of the initial energy 
$E_{51} = E/(10^{51} \rm{erg})$ deposited in the jet, that is $\eta_{10} = 
(10 \% \, E)/(10^{50} \rm {erg})$. It means that in the Hubble time, the 
energetic contribution per unit volume inside the whole universe given by all 
the GRBs is equal to $10^{-21.0} (h^3 \theta_{-1 \rm{j}}^{-2} \eta_{10}) \; 
\rm {erg/cm^3}$. 

To derive the spectrum of GRBs and to compare it with the one of cosmic rays 
out of our Galaxy, we need to calculate the normalization factor $N$ in $N \; 
{E \overwithdelims () E_0}^{-2} d E$, where $N$ is expressed in $\rm {[GeV^{-1} 
\; cm^{-2} \; s^{-1} \; sr^{-1}]}$. 

$N$ is obtained directly from the integration of this power law, remembering 
that the result of this integral is equal to the energy per volume produced 
by all the GRBs: 

\begin{equation}
\int^{E_2}_{E_1} {\biggl[ {{4 \pi} \over c} \; N {E \overwithdelims () 
E_0}^{-2} E  d E \biggr]} = 10^{-21.0} (h^3 \theta_{-1 \rm{j}}^{-2} \eta_{10}). 
\end{equation}
The corresponding value is 

\begin{equation}
N \simeq 10^{-8.8} (h^3 \theta_{-1 \rm{j}}^{-2} 
\eta_{10}) \; {\rm {GeV^{-1} \; cm^{-2} \; s^{-1} \; sr^{-1}}}. 
\end{equation}

\begin{figure}[h]
\vspace{6.8cm}
\includegraphics{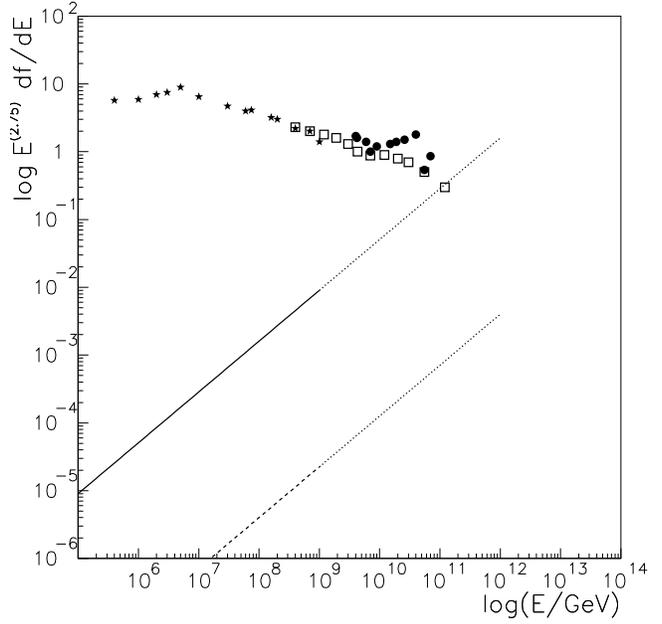} 
\vspace{1.5cm}
\caption[]{\label{cosray_out} 
Comparison between the extragalactic GRB contribution for the case 
a) (solid line) and the case b) (dashed line) and the all 
particle cosmic ray spectrum, expressed in $\rm {[GeV^{-1} \; 
cm^{-2} \; s^{-1} \; sr^{-1}]}$. The dotted lines show that it not 
possible to have any contribution beyond $10^{18}$ eV, because of 
the interactions with the microwave background. The stars represent 
the cosmic ray data from Akeno experiment (Nagano et al. 1984), the 
open squares are the Fly's Eye data (Baltrusaitis et al. 1985) 
and the full circles are the AGASA data (Yoshida et al. 1995).} 
\end{figure}

\item [b)]

We assumed that the total energy discharged by each GRB in the ISM for hadrons 
as well as neutrinos is proportional to its own fluence in the $\gamma$ band. 
To obtain the energetic contribution per unit volume inside the whole universe 
given by all the GRBs in the Hubble time, we integrated the Luminosity Function 
distribution and had $10^{-23.6} (h^3 \theta_{-1 \rm{j}}^{-2} \eta_{10}) \; \rm 
{erg/cm^3}$. 

The corresponding normalization factor $N$ in the curve to plot $N \; 
{E \overwithdelims () E_0}^{-2} d E$ is 

\begin{equation}
N \simeq 10^{-11.4} (h^3 \theta_{-1 \rm{j}}^{-2} \eta_{10}) \; 
{\rm {GeV^{-1} \; cm^{-2} \; s^{-1} \; sr^{-1}}}. 
\end{equation}

\end{description}

In Fig.~\ref{cosray_out} we plot the all particle energy spectrum as 
measured by different ground-based experiments (Biermann $\&$ Wiebel-Sooth 
1999), the spectrum in the case in which each GRB gives the same contribution 
to CR's (solid line) and the spectrum corresponding to a contribution from 
each GRB proportional to its own fluence (dashed 
line). The dotted lines represent unlikely contributions because beyond 
$10^{18}$ eV the interactions with the cosmological microwave background are 
relevant and they make the curves much flatter, and therefore much lower. 
From this graphic it is clear that {\it in our model} GRBs do not give any 
energetic contribution to the extragalactic cosmic ray spectrum. 

\subsection{Galactic contribution}

To calculate the contribution from GRBs to the Galactic cosmic ray spectrum 
it is necessary to know the GRBs production rate in our Galaxy. We considered 
that the ratio between the in/out GRB rates is equal to the ratio of the 
infrared (IR) luminosity inside and outside our Galaxy. In our Galaxy this 
luminosity is $L_{\rm{IR}} \simeq 10^{10} L_{\odot} \simeq 10^{43.6} 
\rm{erg/s}$ at $60 \mu{\rm m}$. Using the equation (1) of Malkan $\&$ 
Stecker (1998), we have an extragalactic IR luminosity $L_{\rm{IR}} \simeq 
10^{44.6} h^3 \rm{erg} \, {(100 \, \rm{Mpc^3})}^{-1}$. This means that the 
Galactic GRB rate is equal to $10^{-6.4} (h^3 \theta_{-1 \rm{j}}^{-2} 
\eta_{10}) \; {\rm {GRBs \; per \; year}}$. 

\begin{description}

\item[a)]

Following the same procedure used for the extragalactic case, we calculated 
first the case in which each GRB gives the same contribution to the cosmic ray 
spectrum. We obtained a diffuse density energy of GRBs in our Galaxy equal to  
$10^{-16.0} (h^3 \theta_{-1 \rm{j}}^{-2} \eta_{10}) \; \rm {erg/cm^3}$.  

In our Galaxy, cosmic rays have a time scale to escape with an energy 
dependence that goes as $E^{-1/3}$ (Biermann 1995, Biermann et al. 1995). 
This term has to be taken in account to calculate the total spectrum of the 
primary cosmic rays inside our Galaxy. It means that it is necessary to 
multiply the injection spectrum $E^{-2}$ times the leakage term $E^{-1/3}$ to 
obtain the final plot of the contribution from GRBs to the Galactic cosmic ray 
spectrum: $N \; {E \overwithdelims () E_0}^{-7/3} d E$. The normalization 
factor is obtained using the same procedure of the last section: 

\begin{equation}
\int^{E_2}_{E_1} {\biggl[ {{4 \pi} \over c} \; N {E \overwithdelims () 
E_0}^{-2} E  d E \biggr]} = 10^{-16.0} (h^3 \theta_{-1 \rm{j}}^{-2} \eta_{10}). 
\end{equation}
The corresponding value is 

\begin{equation}
N \simeq 10^{-3.9} (h^3 \theta_{-1 \rm{j}}^{-2} 
\eta_{10}) \; {\rm {GeV^{-1} \; cm^{-2} \; s^{-1} \; sr^{-1}}}. 
\end{equation}

\begin{figure}[h]
\vspace{6.8cm}
\includegraphics{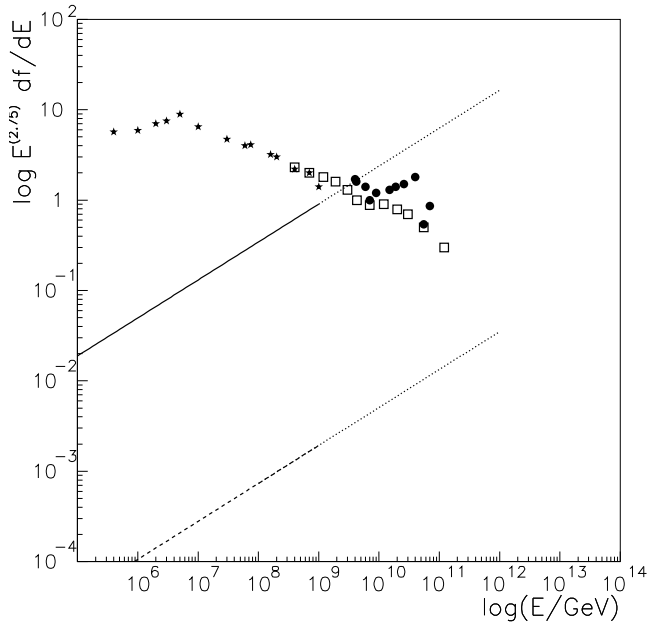} 
\vspace{1.5cm}
\caption[]{\label{cosray_in} 
Comparison between the Galactic GRB contribution for the case 
a) (solid line) and the case b) (dashed line) and the all 
particle cosmic ray spectrum, expressed in $\rm {[GeV^{-1} 
\; cm^{-2} \; s^{-1} \; sr^{-1}]}$. The dotted lines and the 
points are the same of Fig.~\ref{cosray_out}.} 
\end{figure}

\item[b)]

For the second case we assume that each GRB contribution is proportional to 
its own fluence in the $\gamma$-band, and integrating the luminosity function 
distribution we had an energy density in our Galaxy equal to $10^{-18.7} (h^3 
\theta_{-1 \rm{j}}^{-2} \eta_{10}) \; \rm {erg/cm^3}$. 

The corresponding normalization factor $N$ in the spectrum $N \; {E 
\overwithdelims () E_0}^{-7/3} d E$ is 

\begin{equation}
N \simeq 10^{-6.5} (h^3 \theta_{-1 \rm{j}}^{-2} \eta_{10}) 
\; {\rm {GeV^{-1} \; cm^{-2} \; s^{-1} \; sr^{-1}}}. 
\end{equation}

\end{description}

In Fig.~\ref{cosray_in} we compared the all particle energy spectrum as 
measured by different ground-based experiment (Biermann $\&$ Wiebel-Sooth 
1999), with the spectrum from GRBs in the case that each of them gives the 
same contribution (solid line) and with the one in which the contribution is 
proportional to the fluence (dashed line). The dotted lines beyond $10^{18} 
\, {\rm eV}$ show where interactions with the microwave background may become 
relevant. Since massive star formation is highest in the Galactic central 
region, any contribution is limited to energies for which the Larmor radius 
becomes as large as the Galactic disk, and the AGASA data suggest only a small 
anisotropy at the Galactic center. But even if from an energetic point of 
view, GRBs could give a contribution at these high energies, this is ruled out 
considering that the temporal interval between two GRBs in our Galaxy is 
larger than the leakage time of cosmic rays at these energies. Therefore, 
{\it in our model }also in our Galaxy GRBs cannot give any contribution to 
the cosmic ray spectrum. 

\section{Neutrino production}

We also would like to probe the energetic contribution of GRBs to the 
neutrino flux. Both the very high energy (VHE) neutrinos, with energies 
in the range $10^{10} \div 10^{17}$ eV, and the ultra high energy (UHE) 
neutrinos, with energies $E \ge 10^{17}$ eV originate in the interactions 
of protons with photons, through the reactions $p \gamma \to n \pi^{+}$, 
$\pi^{+} \to \mu^{+} \bar{\nu_{\rm \mu}}$, $\mu^{+} \to e^{+} \nu_{\rm \mu} 
\bar{\nu_{\rm e}}$. The efficiency for the neutrino production depends on the 
fraction of proton energy converted into charged pions and on how much 
energy pions and muons keep before decaying. 

Some authors (Waxman $\&$ Bahcall 1997), proposed that GRBs can be associated 
with neutrinos produced in the photohadronic reaction. Pions come from the 
interactions between accelerated protons and gamma rays in the fireball, and 
their decay produces neutrinos and anti-neutrinos together with other 
elementary particles. Waxman $\&$ Bahcall (1997) gave an upper limit for the 
corresponding neutrino flux. 

Other authors (Rachen $\&$ M{\'e}sz{\'a}ros 1998), argued that the upper limit 
obtained in this way is optimistic. In fact, protons emitted in the earliest 
burst do not have enough energy to leave the expansion region because of 
adiabatic losses. Instead neutrons can easily escape and contribute to the 
neutrino flux. 

To check what is the neutrino flux in our model, we calculated the initial 
photon density number in the $p \gamma$ interaction from the synchrotron 
photons in our model. We obtained a number of ${\rm{photons}/ {\rm{cm^3}}}$ 
equal to: 

\begin{eqnarray}
{{L^{\rm{(sh)}}}/ {\biggl ({4 \pi {z \over {4 \gamma_{\rm{sh}}}} 
{{100 \, \rm{KeV}} \over \gamma_{\rm sh}} c} \biggr )}} 
& = & 6.94 \times 10^{16} (E_{51}^{-1/4} \dot M_{-5 \rm j}^{5/4} 
      \times \nonumber \\ 
&   & v_{0.3}^{-5/4}) \theta_{-1 \rm j}^{-2} 
      t^{-7/4}. 
\end{eqnarray}

To obtain the number of hits in the $p \gamma$ collisions, we used a cross 
section $\sigma_{\rm{p \gamma}} = 2.70 \times 10^{-28} \rm{cm^2}$, 
corresponding to the maximum of the curve describing this reaction. This 
implies a number of hits per proton equal to $0.99 \; (E_{51}^{-1/4} \dot 
M_{-5 \rm j}^{5/4} v_{0.3}^{-5/4}) \; \theta_{-1 \rm j}^{-2} t^{-3/4}$. 

\begin{figure}[h]
\vspace{6.8cm}
\includegraphics{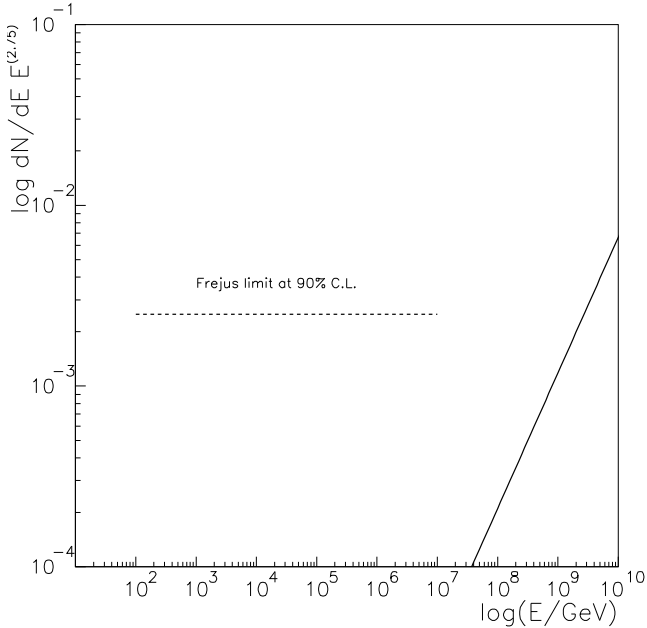} 
\vspace{1.5cm}
\caption[]{\label{fig:neutrino}
The GRB contribution to the neutrino spectrum from our model, 
expressed in ${\rm [{GeV^{-1} \; cm^{-2} \; s^{-1} \; sr^{-1}}]}$.
The dashed line represents the measured upper limit to the 
neutrino flux from the Frejus experiment (Rhode et al., 1996). 
Beyond $10^9$ GeV, corrections from the proton and neutron 
interactions with the microwave background can be expected.} 
\end{figure}

To calculate the energy rate of neutrinos in our model we assumed that 
about $20 \%$ of the energy of protons $E_{\rm p} = 10^{50} \, \rm{erg}$ 
goes into neutrinos and that in a first approximation there are no multiple 
hits for the same proton in the $p \gamma$ interaction. Following the same 
procedure used to calculate the extragalactic cosmic ray contribution from 
GRBs, we obtained that the normalization factor corresponding to the spectrum 
$(E/E_0)^{-2}$ in the case in which each GRB gives the same contribution to 
the neutrino spectrum is $N \simeq 1.55 \times 10^{-10} (h^3 \theta_{-1 
\rm{j}}^2 \eta_{10}) \; {\rm {GeV^{-1} \; cm^{-2} \; s^{-1} \; sr^{-1}}}$. 
This spectrum is plotted in Fig.~\ref{fig:neutrino} and it shows that in 
{\it our jet model} GRBs give a contribution to the cosmological neutrino flux 
at a low level. 

\section{Discussion and conclusions}

In the first part of this article we calculated the GRB rate and compared  
the corresponding cumulative distribution in fluence with the observational 
data. There were two main points to decide on: a) which luminosity function 
distribution and b) which star formation rate were the best to reproduce the 
data. 

We checked if in our jet model GRBs were standard candles. But we did not 
obtain any good fit, therefore we tried a power law for the luminosity 
distribution, $\phi(f) \propto  f^{-\beta}$. Together with this function we 
used the SFR from Miyaji et al. (1998), in which the SFR grows linearly 
from $z=0$ up to about $z=2$ and after this redshift is flat up to $z=6$. 
There was not a good agreement between the distribution in fluence we obtained 
and the corrected data by Petrosian for the 4B BATSE catalogue. Therefore we 
used the SFR model from Madau et al. (1996), in which the rate follows the same 
behavior of Miyaji's up to a redshift $z^{\star}$, and after this redshift 
the SFR begins to decrease. There are still some uncertainties about the shape 
of this second part of the distribution, so we considered different curves 
with different slopes (depending on a constant $a_2$) for this decreasing 
part. We have only two free parameters that we can change to fit the data, 
the power law index $\beta$, and the exponential index $a_2$. The redshift 
$z^{\star}$, and the upper limit $f_{\star \rm{b}}$ of the interval in which 
the LF is defined are not really free parameter because their range is limited 
by the observations. We can reproduce the 4B BATSE corrected data by Petrosian 
using the following values: redshift $z^{\star}=1.7$, $\beta=1.55$, $f_{\star 
\rm{b}}=2.2 \times 10^{-4}$ and $a_2$ in the range [0.8,1.3]. These 
parameters are remarkably constrained. Thus, given a final model for GRBs 
and a cosmological model, we may be able to derive strong limits on 
cosmological parameters. 

The key point of the second part of the article is the calculation of the GRB 
rate inside and outside our Galaxy. The results we obtained depends mainly on 
three other parameters, the value of the Hubble constant $H_{\rm o}$, the 
opening angle of the jet $\theta_{\rm j}$, and the power law index $\alpha$ we 
assumed for the electron distribution in our model (see Eq.~\ref{plato}). 

The choice of $\alpha=2$ has been done in the first version of our model, 
where even if we simplified in many places the physics used, we obtained a 
good agreement with the data. Any possible small changes in this interval 
will influence the results obtained in this work in a marginal way, because 
the dependence on $\alpha$ in the expression of $N$ is not strong. However, 
putting $\alpha = 3$, which is not suggested by the data, would strongly 
influence our results. 

On the other hand, changes in $H_{\rm o}$ and $\theta_{\rm j}$, because of 
the strong dependence in $N$ will influence the results. It is interesting 
to note that both, a lower value of the Hubble constant and a smaller opening 
angle of the jet, go into the direction of decreasing $N$. But these changes 
cannot influence our results because the interactions of cosmic rays with 
the microwave background and the large difference between the Galactic GRBs 
and the leakage time of CR's rule out that {\it in our jet model} GRBs can give 
any contribution beyond $10^{18}$ eV. 

In the calculation of the contribution from GRBs to the neutrino flux, it is 
important to define the parameters that characterize the neutrino production. 
The number of hits in the $p \gamma$ collisions has been obtained considering 
only the major photohadronic interaction channel, the one that gives a 
single pion. At higher energies it is possible to have the channels for the 
production of multi-pions, $p \gamma \to n 2 \pi^{+} \pi^{-}$ and $p \gamma 
\to n 3 \pi^{+} 2 \pi^{-}$, but as the energy increases the corresponding 
cross sections decrease. We used only the first channel with the highest 
cross section, because the energies involved are not high enough to require 
secondary channels. 

In the context of our {\it jet-disk symbiosis model} for GRBs, there is only 
one way to make the extragalactic contribution to CRs significant at the 
highest energies, and that is to drastically increase the CR energies per GRB 
deposited. We consider this implausible in the context of our model because 
of energy conservation requirements. We use $10 \%$ of the entire energy 
available, and so the CR contribution may be increased over our simple 
calculation by a factor of a few, but not more. This implies that the 
contributions given by each GRB to CR and neutrino flux cannot be much 
bigger than the energy emitted in the $\gamma$-ray band. 

Using a relatively small set of parameters, the jet-disk symbiosis model 
applied to GRBs, a tested star formation rate and the fundamental physics 
of the photohadronic interactions we arrive at the conclusion that GRBs are 
not able to give any significant contribution to the high energy cosmic ray 
spectrum both inside and outside our Galaxy and predict only a low flux 
of neutrinos. 

A main conclusion of this work is that fitting the corrected fluence 
distribution of GRBs with the jet-model is well possible. The fit allows 
strong constraints of the star formation rate as a function of redshift. 

\vspace{0.3cm}

\begin{acknowledgements}  
GP thanks S. S. Larsen and E. Ros for useful discussions. We are 
grateful to V. Petrosian for the 4B corrected data he gave us, and for 
discussions. We want also to thank our referee R.A.M.J. Wijers for the 
helpful advice and comments that he gave us to improve our article. PLB 
wishes to thank T. Piran for extensive discussions of GRBs. GP is supported 
by a DESY grant 05 3BN62A 8. HF is supported by a DFG grant 358/1-1\&2.
\end{acknowledgements}

\end{document}